\documentclass[]{spie}  

\usepackage{amsmath,amsfonts,amssymb}
\usepackage{graphicx}
\usepackage[colorlinks=true, allcolors=blue]{hyperref}

\title{Fully broadband vAPP coronagraphs enabling polarimetric high contrast imaging }

\author[a]{Steven P. Bos}
\author[a]{David S. Doelman}
\author[a]{Jos de Boer}
\author[a]{Emiel H. Por}
\author[b]{Barnaby Norris}
\author[c]{Michael J. Escuti}
\author[a]{Frans Snik}
\affil[a]{Leiden Observatory, Leiden University, P.O. Box 9513, 2300 RA Leiden, The Netherlands}
\affil[b]{Sydney Institute for Astronomy (SIfA), Institute for Photonics and Optical Science (IPOS), School of Physics, University of Sydney, NSW2006, Australia}
\affil[c]{Department of Electrical and Computer Engineering, North Carolina State University, Raleigh, NC 27695, USA}

\authorinfo{email: stevenbos@strw.leidenuniv.nl}

\begin{document} 
\maketitle

\begin{abstract}
We present designs for fully achromatic vector Apodizing Phase Plate (vAPP) coronagraphs, that implement low polarization leakage solutions and achromatic beam-splitting, enabling observations in broadband filters. The vAPP is a pupil plane optic, inducing the phase through the inherently achromatic geometric phase. We discuss various implementations of the broadband vAPP and set requirements on all the components of the broadband vAPP coronagraph to ensure that the leakage terms do not limit a raw contrast of $10^{-5}$. Furthermore, we discuss superachromatic QWPs based of liquid crystals or quartz/MgF2 combinations, and several polarizer choices. As the implementation of the (broadband) vAPP coronagraph is fully based on polarization techniques, it can easily be extended to furnish polarimetry by adding another QWP before the coronagraph optic, which further enhances the contrast between the star and a polarized companion in reflected light. We outline several polarimetric vAPP system designs that could be easily implemented in existing instruments, e.g. SPHERE and SCExAO.
\end{abstract}


\section{INTRODUCTION}\label{sec:intro}
\begin{figure}
\begin{center}
\begin{tabular}{c}
\includegraphics[width=0.75\textwidth]{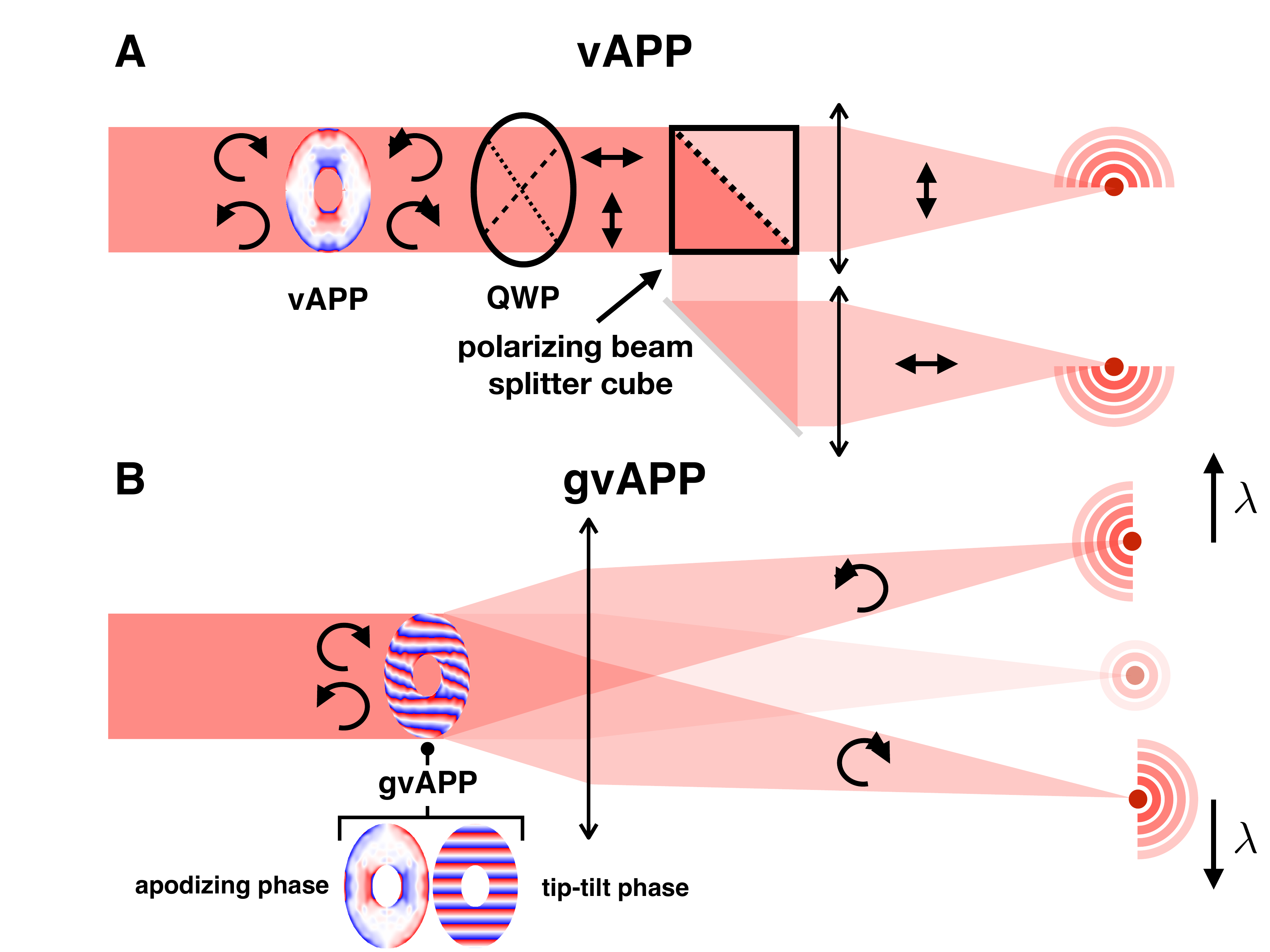}
\end{tabular}
\end{center}
\caption 
{ \label{fig:figure_1} Sketches of (\textbf{A}) the vector Apodizing Phase Plate (vAPP) and (\textbf{B}) the grating vAPP (gvAPP).} 
\end{figure}

\indent High contrast imaging instruments investigate the circumstellar region direct around the star to search for exoplanets and signs of planet formation. One of the most exciting goals is the direct imaging and characterization of Earth-like exoplanets to search for signs of life. This is incredibly hard as an Earth-like exoplanet around a Sun-like star at 50 pc will have an angular separation of $\sim50$ mas and in the visible a contrast of $\sim10^{-10}$. This requires suppression of the diffracted starlight, many orders of magnitude brighter than the emitted light or reflected starlight from the exoplanet, with a coronagraph. Adaptive optics (AO) systems on ground-based systems are unable to give a perfect correction to wavefront distorted by the atmosphere, and therefore the raw contrast achievable by coronagraphs at small inner working angles (IWA) is fundamentally limited to $\sim10^{-5}$\cite{guyon2012elts}. The raw contrast will be further degraded by quasi-static aberrations originating from within the telescope and instrument, e.g. due to changes in temperature and gravitational vector. To reach the contrast required for Earth-like exoplanets, diversity between the light from the star and planet has to be utilised. Techniques such as angular differential imaging\cite{marois2006angular} (ADI) and spectral differential imaging\cite{sparks2002imaging} (SDI) are very successful, but have a reduced effectiveness at small IWA when the diversities they rely on diminish in these regions. High-dispersion ($R \sim 100.000$) spectroscopy\cite{sparks2002imaging} (HDS) and polarimetric differential imaging\cite{snik2013astronomical} (PDI) are techniques capable of increasing contrast at small IWA.\\
\\
\indent The Apodizing Phase Plate \cite{codona2004imaging}$^,$ \cite{codona2006high}$^,$ \cite{kenworthy2007first} (APP) is a transmissive optic that sculpts a $180^{\circ}$ dark hole in the point spread function (PSF) from 2 to 7 $\lambda / D$ at one side of the star and generally gives a $10^{-4}$ contrast. The APP applies a phase in the pupil plane by optical path differences (OPD) caused by the varying height of a diamond turned zinc selenide plate. Being a pupil plane coronagraph makes the APP robust against tip-tilt errors and resolved stars\cite{guyon2006theoretical}, problems that focal plane coronagraphs limit in performance. The diamond turning process is the limiting factor for the APP as it requires smoothly varying phase patterns that: 1) will not give the optimal contrast and strehl performance, 2) only allow for $180^{\circ}$ dark holes requiringtwo observations to investigate the full circumstellar region, and 3) is very chromatic due to the wavelength dependence of the OPD.\\
\\
\indent The vector APP\cite{snik2012vector} (vAPP) solves these problems by inducing the phase through the geometric phase\cite{pancharatnam1956generalized}$^,$ \cite{berry1987adiabatic} on the circular polarization states. The vAPP is an liquid crystal plate that is an half-wave retarder with a spatially varying fast axis. The acquired phase $\phi$ by the $\pm$circular polarization states depends locally on the fast axis angle $\theta$, and is given by $\phi = \pm 2 \theta$. The geometric phase is inherently achromatic, but the efficiency of the light that acquires this phase is determined by retardance of the liquid crystal layer. Any retardance offsets from half-wave will result in leakage that did not acquire the desired phase. A direct-writing systems\cite{miskiewicz2014direct} allows for extreme fast axis orientation patterns that are required for optimally designed phase designs for arbitrary telescope apertures\cite{por2017optimal} and $360^{\circ}$ dark holes\cite{otten2014performance}. Self aligning liquid crystal layers \cite{komanduri2012multi}$^,$\cite{komanduri2013multi} enable wave plates close to half-wave over broad wavelength ranges. The vAPP yields two PSFs with opposite circular polarization and opposite dark holes that need to be separated. The first iteration\cite{otten2014performance} of the vAPP separated the two PSFs by adding a quarter-wave plate (QWP) and a polarizer (e.g. a Wollaston prism or a polarizing beam splitter cube), see \autoref{fig:figure_1} A. For the achievable raw contrast not be limited by leakage terms sets tight tolerances on the retardance and fast axis offsets of the vAPP and QWP, and the extinction ratio of the polarizer. Leakage due to retardance offsets of the half-wave liquid crystal layer imprints a copy of the non-coronagraphic in the dark holes, while leakage of similar and fast axis orientation offsets by the QWP and polarizer mixes the two coronagraphic PSFs. Another way of splitting the two PSFs and simultaneously separating the coronagraphic PSFs from the leakage PSF, is by adding a tip-tilt phase ramp to the apodizing phase pattern itself, as is shown in \autoref{fig:figure_1} B. This is known as a polarization grating\cite{oh2008achromatic} (PG), that separates the opposite circular polarization states, and these coronagraphs are therefore known as grating vAPPs (gvAPP). Due to wavelength smearing by the grating, the gvAPPs are mainly suitable for narrowband imaging and low resolution integral field spectrographs (IFS) such as CHARIS\cite{peters2013optical}$^,$\cite{brandt2014charis} and SPHERE-IFS\cite{claudi2008sphere}$^,$ \cite{mesa2015performance}. The gvAPP has already been put on sky with several instruments, e.g. SCExAO\cite{doelman2017patterned}, MagAO/Clio2\cite{otten2017sky} and LEXI\cite{haffert2018lexi}.\\
\\
\indent The gvAPP has excellent broadband performance, but is not suitable for broadband imaging (polarimetry) or as a combination with a high resolution IFS\cite{snellen2015combining}. For this we have to turn back to PSF splitting with QWPs and polarizers. In this proceeding we will explore how fully broadband vAPPs and polarimetric vAPPs can be implemented in existing high contrast imaging systems such as SPHERE\cite{beuzit2008sphere} and SCExAO\cite{jovanovic2015subaru}. 
\section{Fully broadband vAPPs}\label{sec:broadband_vAPP}
\begin{figure}
\begin{center}
\begin{tabular}{c}
\includegraphics[width=0.75\textwidth]{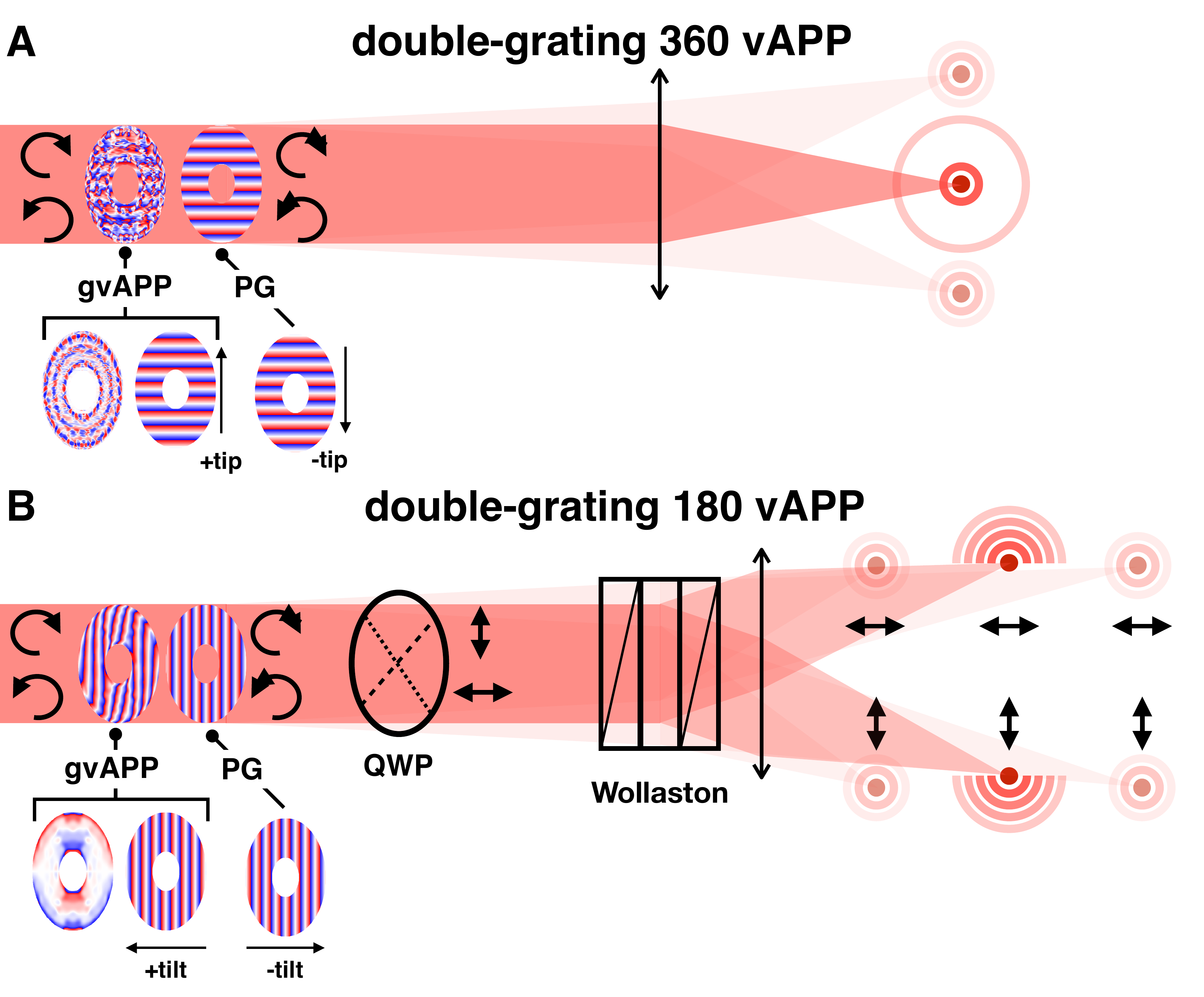}
\end{tabular}
\end{center}
\caption 
{ \label{fig:figure_3} Double-grating implementations of (\textbf{A}) the $360^{\circ}$ vAPP and (\textbf{B}) the $180^{\circ}$ vAPP by combining a gvAPP with an additional polarization grating (PG) with opposite tip-tilt.}  
\end{figure}

Fully broadband vAPPs that do not suffer from wavelength smearing have different implementations depending on the symmetric (e.g. $360^{\circ}$ vAPP) or anti-symmetric (e.g. $180^{\circ}$ vAPP) coronagraphic PSFs. Symmetric vAPPs can benefit from a second PG with an opposite tip-tilt after a gvAPP (the double-grating vAPP\cite{doelman2017patterned}) that folds back the PSFs back on top of each other, see \autoref{fig:figure_3} A. This has the added benefit that leakage by the vAPP is greatly reduced by the combination of the two gratings and has already be implemented\cite{doelman2017patterned} for LMIRcam\cite{skrutskie2010large} on the LBT telescope.\\
\\
\indent Asymmetric vAPPs can also benefit from the leakage reduction by the double-grating, see \autoref{fig:figure_3} B, but still need another method to separate the PSFs, otherwise the dark hole of one PSF will be filled by the bright side of the other. To separate the PSFs we go back to the first iteration\cite{otten2014performance} of the vAPP, i.e. add an QWP and polarizer downstream of the vAPP, see \autoref{fig:figure_1} A and \autoref{fig:figure_3} B. This sets tight tolerances on the retardance and fast axis offsets of the vAPP and QWP, and the extinction ratio (ER) of the polarizer. Leakage from the vAPP will imprint a copy of the non-coronagraphic PSF and is determined by retardance offsets ($\Delta \delta$) from half-wave. Offsets of the fast axis orientation change the imprinted phase and are not considered in this work. The intensity of the leakage term ($I_{HWP}$) is given by:
\begin{equation}\label{eq:leakage_HWP}
I_{HWP} = \sin(\frac{1}{2} \Delta \delta)^2. 
\end{equation}
Leakage by the QWP and polarizer will imprint a copy of the one coronagraphic PSF onto the other. This sets much stricter tolerances on these elements as the bright side of the one coronagraphic PSF will leak into the dark hole of the other. This leakage is stronger than leakage by the non-coronagraphic PSF. Leakage by the QWP is not only dependent on retardance offsets ($\Delta \delta$), but also on fast axis orientation offsets ($\Delta \theta$) w.r.t. polarizer. The total leakage ($I_{QWP}$) (ignoring an irrelevant cross term) is given by:
\begin{equation}\label{eq:leakage_QWP}
I_{QWP} \approx \sin(\frac{1}{2} \Delta \delta)^2 + \sin(\Delta \theta)^2.
\end{equation}
Leakage by the polarizer is simply given by the extinction ratio.\\
\\
\indent The vAPP and QWP can be optically contacted to form one optic, the polarizer can be further downstream in the system. It is critical to ensure that the QWP maps the circular polarization states (with the imprinted phase patterns) to the linear polarization states that are separated by the polarizer, i.e. the QWP and polarizer need to be carefully aligned.\\
\\
\indent Requirements on the components of the $180^{\circ}$ vAPP for a raw contrast of $<10^{-5}$ are given in \autoref{tab:table_1}.  The leakage requirements are determined by the ratio of the intensities at the IWA between respectively the coronagraphic PSF and non-coronagraphic PSF (for the HWP), and the opposite coronagraphic PSFs (for the QWP and polarizer). \autoref{eq:leakage_HWP} and \autoref{eq:leakage_QWP} translate this to the requirements on the retardance and fast axis orientation.
\begin{table}
\begin{center}
\begin{tabular}{ | l | l | l |}
\hline
  Requirements & Leakage & Retardance and fast axis angle \\ \hline
  HWP & $< 1 \cdot 10^{-3}$ & $\Delta \delta < 3.6^{\circ}$ \\ \hline
  QWP & $< 5 \cdot 10^{-5}$ & $\Delta \delta < 0.8^{\circ}$, $\Delta \theta < 0.4^{\circ} $ \\ \hline
  polarizer & $< 5\cdot10^{-5}$ & ER. $ < 20.000:1$ \\ \hline
\end{tabular} 
\caption{\label{tab:table_1} Requirements for $180^{\circ}$ vAPPs with a raw contrast $< 10^{-5}$.} 
\end{center}
\end{table}
The requirements of the $360^{\circ}$ vAPP are slightly relaxed due to the (generally) larger IWA and therefore the more favourable ratio between the intensities of the coronagraphic and non-coronagraphic PSF. Requirements on the QWP and polarizer are absent as the double-grating symmetric vAPPs do not need these components.\\
\begin{table}
\begin{center}
\begin{tabular}{ | l | l | l |}
\hline
  Requirements & Leakage & Retardance and fast axis angle \\ \hline
  HWP & $< 2.5 \cdot 10^{-3}$ & $\Delta \delta < 5.7^{\circ}$ \\ \hline
  QWP &  & \\ \hline
  polarizer &  & \\ \hline
\end{tabular} 
\caption{\label{tab:table_2} Requirements for $360^{\circ}$ vAPPs with a raw contrast $< 10^{-5}$. } 
\end{center}
\end{table}

\subsection{Components}
\autoref{tab:table_1} and \autoref{tab:table_2} set the requirements on the components for a raw contrast of $<10^{-5}$. For a broadband vAPP this means that over a sufficient wavelength range the components need to meet these requirements. The vAPP is a stack of self-aligning liquid crystal layers that have a half-wave retardance with a spatially varying fast axis orientation. Half-wave liquid crystal retarders have been fabricated with leakage of $\sim10^{-2}$ over broad wavelength ranges in the optical\cite{komanduri2012multi} (400-750 nm), near infrared\cite{doelman2018first} (900-2400 nm) and infrared\cite{otten2017sky} (2000-4500 nm). These retarders meet, combined with the double-grating technique, the leakage requirements. Furthermore, the liquid crystals have also been shown to operate in cryogenic environments.\cite{otten2017sky} \\
\\
\indent The requirements for the QWP are much tighter and cannot benefit from the double-grating technique. However, there is a broader choice of materials available as the fast axis orientation constant within the optic. In the NIR an off-the-shelf superachromatic QWP (Pancharatnam configuration of three MgF2/quartz plates) provided by Thorlabs already meets the requirements over a broad (900 - 2700 nm) wavelength range. Similar wave plates in the visible and infrared off-the-shelf products that meet the requirements are currently not available. Therefore, custom QWP designs have to be pursued that might employ more MgF2/quartz plates. For MgF2/quartz retarders further investigation is needed to identify the impact of fringing, their thermal properties and wavefront quality.  Liquid crystal QWPs\cite{komanduri2012multi} are available as well that can operate in the required wavelength ranges and are less prone to fringing problems, thermal expansion and reach a satisfactory wavefront error (WFE). Further research is required to investigate if such wave plates can be fabricated that reach a satisfactory retardance and fast axis orientation performance.\\
\\
\indent The choice of polarizer mainly depends on the optical system, but polarizing beam splitters are favoured for the sake of photon efficiency. Wollaston prism are an ideal choice: high extinction ratio ($\sim$100.000:1) and achromatic. Polarizing beam splitter cubes/plates with dielectric and/or metallic coatings  (wire grid cubes suffer from high WFE in the reflected beam) combined with additional (wire grid) polarizers in the two beams to increase the polarization purity, can also reach a sufficient broadband performance. 
\subsection{Implementations for existing instruments}
\subsubsection{SPHERE-IRDIS}
\begin{figure}
\begin{center}
\begin{tabular}{c}
\includegraphics[width=0.75\textwidth]{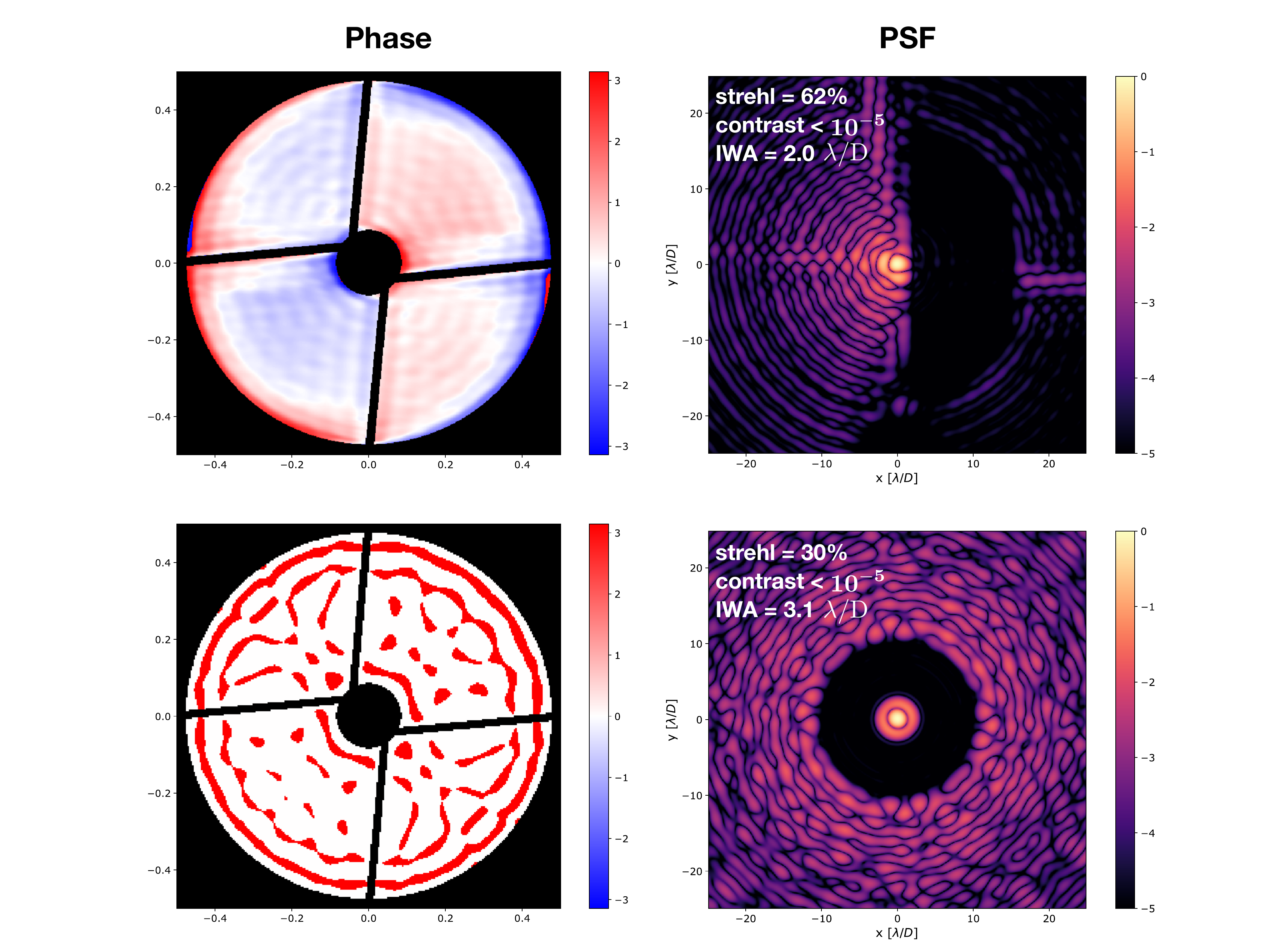}
\end{tabular}
\end{center}
\caption 
{ \label{fig:SPHERE_designs} Preliminary $180^{\circ}$ (top row) and $360^{\circ}$ (bottom row) vAPP designs for SPHERE.} 
\end{figure}

Within the high contrast imaging instrument SPHERE\cite{beuzit2008sphere} and behind the extreme AO system SAXO\cite{fusco2006high} sits the NIR imaging arm IRDIS\cite{dohlen2008infra}. Within this arm there are two pupil planes (pre-apodizer and Lyot) available where a broadband vAPP can be installed. The pre-apodizer wheel is preferable, because downstream there can be a focal plane mask installed that prevents saturation effects with long integration times. A double-grating $360^{\circ}$ vAPP (see \autoref{fig:SPHERE_designs} for the preliminary design) with a liquid crystal recipe similar to the SCExAO gvAPP\cite{doelman2018first} can easily be installed. A $180^{\circ}$ vAPP (\autoref{fig:SPHERE_designs}) requires an additional optically contacted (to the liquid crystal plate) QWP and the existing wire grid polarizers\cite{langlois2014high} (if their ER is sufficient). To increase the photon efficiency, an upgrade of the existing beam splitter plate to an polarizing beam splitter plate would be favourable. A focal plane mask for the $180^{\circ}$ would require a combination of a neutral density filter and a polarizer as the two corongraphic PSFs still overlap.

\subsubsection{SCExAO}
\begin{figure}
\begin{center}
\begin{tabular}{c}
\includegraphics[width=0.75\textwidth]{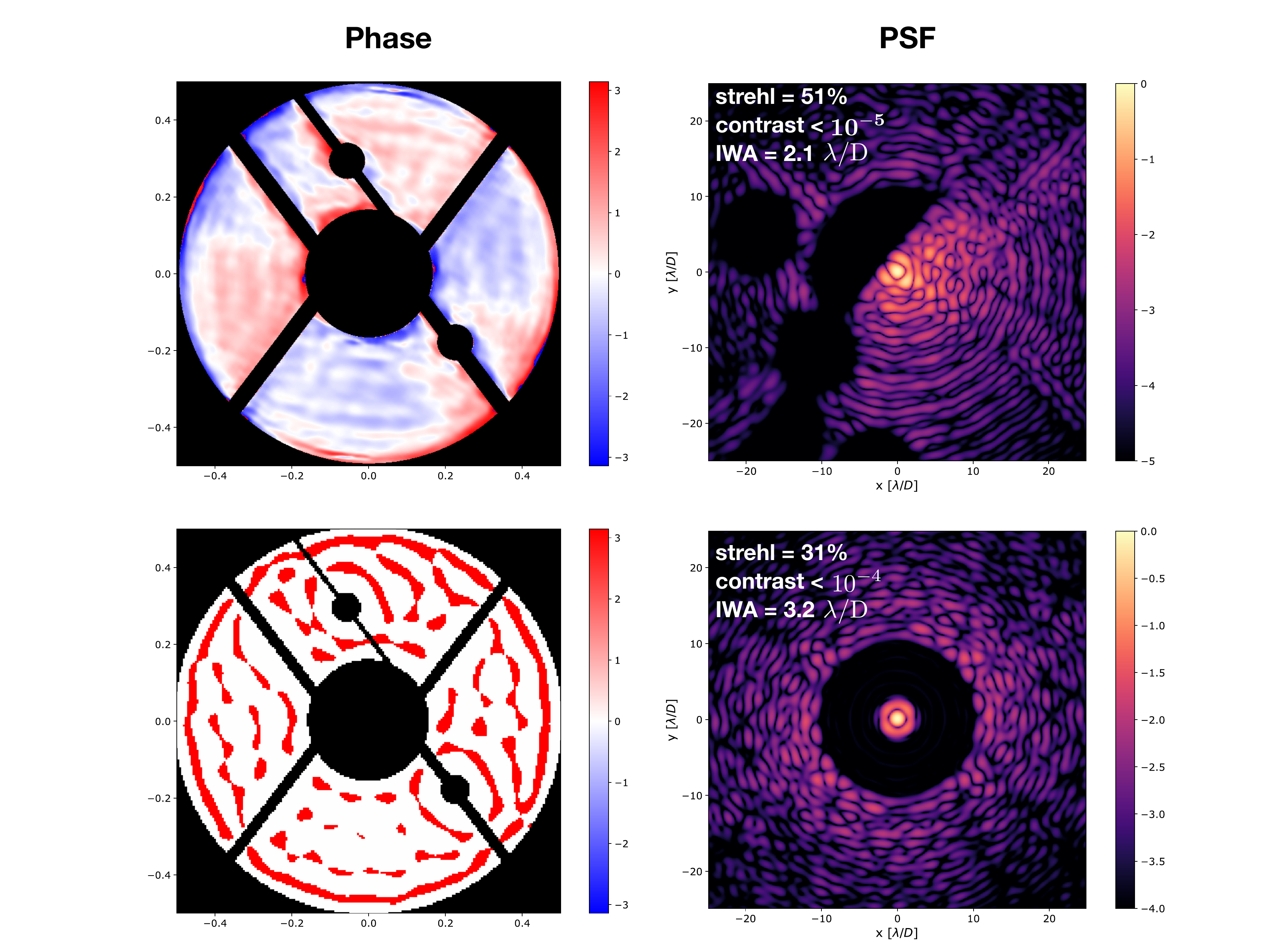}
\end{tabular}
\end{center}
\caption 
{ \label{fig:SCExAO_designs} Preliminary $180^{\circ}$\cite{doelman2017patterned} (top row) and $360^{\circ}$\cite{por2017optimal} (bottom row) vAPP designs for SCExAO. Note that the design for the $180^{\circ}$ vAPP has been used as well for the SCExAO gvAPP\cite{doelman2017patterned}. The SCExAO vAPP has a more complicated dark hole structure to support phase diversity spots and photometry on the leakage PSF, this might change for the final vAPP design.} 
\end{figure}

Similar designs can also be put in SCExAO\cite{jovanovic2015subaru}, see \autoref{fig:SCExAO_designs} for the phase designs. They could be installed in the pupil plane where the current SCExAO gvAPP\cite{doelman2018first}  resides. There will be Wollaston prisms installed\cite{lozi2018scexao} in CHARIS and in front of the CRED-2 camera that can be used for the beam-splitting of the $180^{\circ}$ vAPP. 
\section{Polarimetric vAPPs}\label{sec:polarimetric_vAPP}
\begin{figure}
\begin{center}
\begin{tabular}{c}
\includegraphics[width=0.75\textwidth]{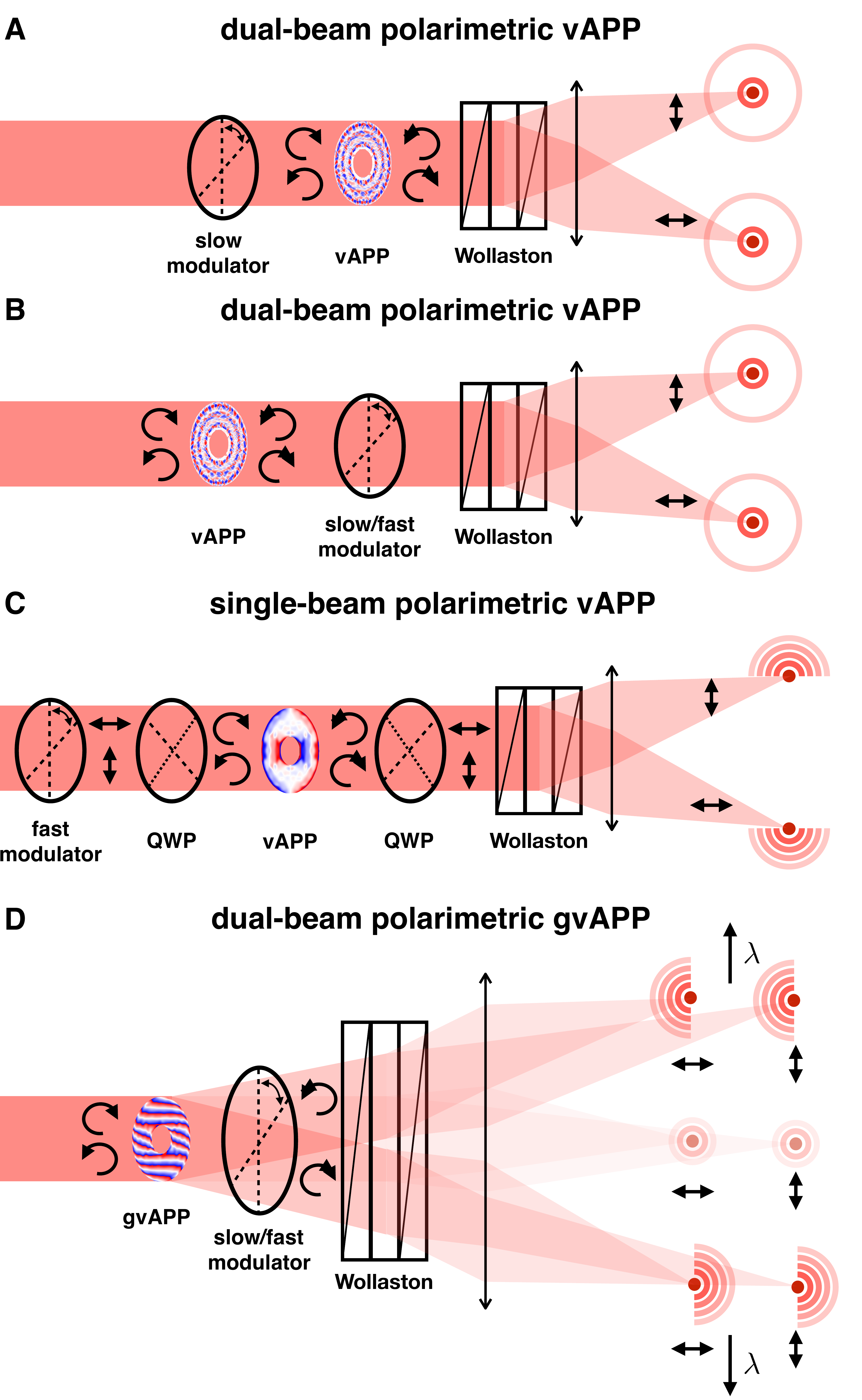}
\end{tabular}
\end{center}
\caption 
{ \label{fig:figure_2} Sketches of several polarimetric vAPP implementations. (\textbf{A}) In case of a slow polarization modulator in front of the coronagraph the solution is an $360^{\circ}$ vAPP implementation that enables dual-beam polarimetry. (\textbf{B}) When the polarization modulator (either fast or slow) is located after the corongraph, again the $360^{\circ}$ vAPP is the design of choice. (\textbf{C}) A fast polarization modulator in front of vAPP allows for a single-beam $180^{\circ}$ vAPP. (\textbf{D}) A gvAPP followed by a (fast or slow) polarization modulator and a Wollaston prism enables dual-beam narrow band imaging polarimetry with the $180^{\circ}$ vAPP. } 
\end{figure}

As the vAPP operates on the polarization nature of light, it can be naturally combined with broadband imaging polarimetry \cite{snik2014combining}. A $360^{\circ}$ vAPP as in \autoref{fig:figure_3} A combined with a beam splitting polarizer (see \autoref{fig:figure_2} A and B) is a dual-beam polarimetric system, as both PSFs measure an orthogonal polarization and can directly be subtracted and added due to identical morphologies of the PSFs. On the other hand, a $180^{\circ}$ vAPP (\autoref{fig:figure_3} B) combined with an upstream QWP is considered to be a single-beam system, see \autoref{fig:figure_2} C, and has to be temporally modulated as the morphologies of the PSFs are different and cannot be subtracted or added directly. Dual-beam polarimetric $180^{\circ}$ vAPP systems have been explored by Snik\cite{snik2014combining} and will not be discussed here due to the complexity of their implementation. Note that vAPPs are generally designed for a specific pupil shape and focal plane layout, therefore they can only be operated in pupil tracking mode.\\
\\
\indent The specific polarimetric implementation of the vAPP is determined by the symmetry (symmetric/anti-symmetric) of the coronagraphic PSF, the position (up- or downstream of the vAPP) and the speed (faster/slower than coherence time atmosphere) of the polarization modulator, and a single- or dual-beam implementation\cite{snik2013astronomical}. Polarimetric vAPPs with asymmetric PSFs are most straightforward implemented as single-beam systems, while symmetric PSFs can simply be operated as dual-beam systems. Single-beam implementations are only possible with fast polarization modulators (such as a ferroelectric liquid crystal (FLC) ) as both polarization states have to be measured within the timescale that the atmosphere can be considered 'frozen', while dual-beam systems measure both polarization states simultaneously and can therefore handle slow polarization modulators (such as a rotating HWP) as well. When the polarization modulator is downstream of the vAPP, but in front of the polarizer, only dual-beam systems are possible. If the modulator is upstream of the vAPP, both single- and dual-beam systems can be implemented.\\  
\\
\indent The requirements for the components set by \autoref{tab:table_1} and \autoref{tab:table_1} are sufficient for polarimetric purposes and have been extensively discussed in \autoref{sec:broadband_vAPP}.\\

\subsection{Implementations for existing instruments}
\subsubsection{SPHERE-IRDIS}
The polarimetric mode of SPHERE-IRDIS\cite{langlois2014high} has been greatly successful in characterizing the scattered NIR light from debris and protoplanetary disks. The polarization modulator of IRDIS is a slowly rotating HWP upstream of the pupil plane where the polarimetric vAPP would be located. The wire grid polarizers are located after the beam splitter plate, that ideally would be upgraded to an polarization beam splitting version. The polarimetric vAPP would therefore be a dual-beam $360^{\circ}$ vAPP, see \autoref{fig:figure_2} A.

\subsubsection{SPHERE-ZIMPOL}
SPHERE-ZIMPOL\cite{thalmann2008sphere}, operating in visible wavelengths, has a fast polarization modulator (FLC) just before a polarizing beam splitter cube. Unfortunately, there is no pupil plane available between the two components. Therefore, the polarimetric vAPP of choice again is a dual-beam $360^{\circ}$ vAPP, see \autoref{fig:figure_2} B.

\subsubsection{SCExAO}
SCExAO will be upgraded with an NIR FLC\cite{lozi2018scexao} and an Wollaston prism. Down stream of the FLC there would be a pupil plane available where a single-beam $180^{\circ}$ vAPP can be installed, as shown in \autoref{fig:figure_2} C. 

\subsubsection{VAMPIRES}
VAMPIRES\cite{norris2015vampires} is an imaging polarimeter in the visible, that recently got upgraded with an FLC \cite{lozi2018scexao}. Between this FLC and the polarizing beam splitter cube is a pupil plane where a single-beam $180^{\circ}$ vAPP can be installed, as in \autoref{fig:figure_2} C. 

\subsubsection{SCExAO gvAPP}
SCExAO already installed a gvAPP in the NIR\cite{doelman2018first}. With the upcoming NIR FLC and Wollaston prism, this can be combined to a narrowband, dual-beam $180^{\circ}$ polarimetric vAPP, as shown in \autoref{fig:figure_2} D. Note that the other implementations described all are broadband imaging polarimetry implementations.

\section{CONCLUSION}\label{sec:conclusion}
We have discussed fully broadband vAPPs that do not suffer from wavelength smearing. We have shown that the symmetry of the coronagraph determines the exact implementation. Symmetric vAPPs can be implemented with the double-grating technique, anti-symmetric vAPPs require an additional QWP and polarizer to split both coronagraphic PSFs.  We have set requirements on the components of the vAPP such that the leakage will not limit the raw contrast of $10^{-5}$. In the NIR all components are currently available. For the visible and IR wavelengths custom QWPs need to be designed. Further investigation is required to get an overview of all the risks. Furthermore, we have have shown that such vAPPs can be installed, with minimal modifications, in the NIR arms of existing high contrast instruments such as SPHERE and SCExAO. Phase designs and the resulting PSFs are presented as well.\\
\indent Furthermore, we have discussed how polarimetry can be combined with the vAPP. We discussed various implementations, that depend on the symmetry of the coronagraphic PSF, the position and speed of the modulator, and single- or dual-beam polarimetry. We outlined how the polarimetric vAPP could be implemented within SPHERE-IRDIS, SPHERE-ZIMPOL, SCExAO and VAMPIRES.

\acknowledgments 
The research of Steven P. Bos, David S. Doelman, Jos de Boer and Frans Snik leading to these results has received funding from the European Research Council under ERC Starting Grant agreement 678194 (FALCONER). This research used the \textbf{hcipy}\cite{por2018hcipy} Python package that provides a common framework for performing optical propagation simulations, meant for high contrast imaging.

\bibliography{report} 

\begin{thebibliography}{10}

\bibitem{guyon2012elts}
Guyon, O., Martinache, F., Cady, E.~J., Belikov, R., Balasubramanian, K.,
  Wilson, D., Clergeon, C.~S., and Mateen, M., ``How elts will acquire the
  first spectra of rocky habitable planets,'' in [{\em Adaptive Optics Systems
  III}{\nolinebreak\hspace{0.1em}]},   {\bf 8447},  84471X, International
  Society for Optics and Photonics (2012).

\bibitem{marois2006angular}
Marois, C., Lafreniere, D., Doyon, R., Macintosh, B., and Nadeau, D., ``Angular
  differential imaging: a powerful high-contrast imaging technique,'' {\em The
  Astrophysical Journal}~{\bf 641}(1),  556 (2006).

\bibitem{sparks2002imaging}
Sparks, W.~B. and Ford, H.~C., ``Imaging spectroscopy for extrasolar planet
  detection,'' {\em The Astrophysical Journal}~{\bf 578}(1),  543 (2002).

\bibitem{snik2013astronomical}
Snik, F. and Keller, C.~U., ``Astronomical polarimetry: polarized views of
  stars and planets,'' in [{\em Planets, Stars and Stellar
  Systems}{\nolinebreak\hspace{0.1em}]},   175--221, Springer (2013).

\bibitem{codona2004imaging}
Codona, J.~L. and Angel, R., ``Imaging extrasolar planets by stellar halo
  suppression in separately corrected color bands,'' {\em The Astrophysical
  Journal Letters}~{\bf 604}(2),  L117 (2004).

\bibitem{codona2006high}
Codona, J., Kenworthy, M., Hinz, P., Angel, J., and Woolf, N., ``A
  high-contrast coronagraph for the mmt using phase apodization: design and
  observations at 5 microns and 2 $\lambda$/d radius,'' in [{\em Ground-based
  and Airborne Instrumentation for Astronomy}{\nolinebreak\hspace{0.1em}]},
  {\bf 6269},  62691N, International Society for Optics and Photonics (2006).

\bibitem{kenworthy2007first}
Kenworthy, M.~A., Codona, J.~L., Hinz, P.~M., Angel, J. R.~P., Heinze, A., and
  Sivanandam, S., ``First on-sky high-contrast imaging with an apodizing phase
  plate,'' {\em The Astrophysical Journal}~{\bf 660}(1),  762 (2007).

\bibitem{guyon2006theoretical}
Guyon, O., Pluzhnik, E., Kuchner, M., Collins, B., and Ridgway, S.,
  ``Theoretical limits on extrasolar terrestrial planet detection with
  coronagraphs,'' {\em The Astrophysical Journal Supplement Series}~{\bf
  167}(1),  81 (2006).

\bibitem{snik2012vector}
Snik, F., Otten, G., Kenworthy, M., Miskiewicz, M., Escuti, M., Packham, C.,
  and Codona, J., ``The vector-app: a broadband apodizing phase plate that
  yields complementary psfs,'' in [{\em Modern Technologies in Space-and
  Ground-based Telescopes and Instrumentation II}{\nolinebreak\hspace{0.1em}]},
    {\bf 8450},  84500M, International Society for Optics and Photonics (2012).

\bibitem{pancharatnam1956generalized}
Pancharatnam, S., ``Generalized theory of interference and its applications,''
  in [{\em Proceedings of the Indian Academy of Sciences-Section
  A}{\nolinebreak\hspace{0.1em}]},   {\bf 44}(6),  398--417, Springer (1956).

\bibitem{berry1987adiabatic}
Berry, M.~V., ``The adiabatic phase and pancharatnam's phase for polarized
  light,'' {\em Journal of Modern Optics}~{\bf 34}(11),  1401--1407 (1987).

\bibitem{miskiewicz2014direct}
Miskiewicz, M.~N. and Escuti, M.~J., ``Direct-writing of complex liquid crystal
  patterns,'' {\em Optics Express}~{\bf 22}(10),  12691--12706 (2014).

\bibitem{por2017optimal}
Por, E.~H., ``Optimal design of apodizing phase plate coronagraphs,'' in [{\em
  Techniques and Instrumentation for Detection of Exoplanets
  VIII}{\nolinebreak\hspace{0.1em}]},   {\bf 10400},  104000V, International
  Society for Optics and Photonics (2017).

\bibitem{otten2014performance}
Otten, G.~P., Snik, F., Kenworthy, M.~A., Miskiewicz, M.~N., and Escuti, M.~J.,
  ``Performance characterization of a broadband vector apodizing phase plate
  coronagraph,'' {\em Optics Express}~{\bf 22}(24),  30287--30314 (2014).

\bibitem{komanduri2012multi}
Komanduri, R.~K., Kim, J., Lawler, K.~F., and Escuti, M.~J., ``Multi-twist
  retarders for broadband polarization transformation,'' in [{\em Emerging
  Liquid Crystal Technologies VII}{\nolinebreak\hspace{0.1em}]},   {\bf 8279},
  82790E, International Society for Optics and Photonics (2012).

\bibitem{komanduri2013multi}
Komanduri, R.~K., Lawler, K.~F., and Escuti, M.~J., ``Multi-twist retarders:
  broadband retardation control using self-aligning reactive liquid crystal
  layers,'' {\em Optics Express}~{\bf 21}(1),  404--420 (2013).

\bibitem{oh2008achromatic}
Oh, C. and Escuti, M.~J., ``Achromatic diffraction from polarization gratings
  with high efficiency,'' {\em Optics letters}~{\bf 33}(20),  2287--2289
  (2008).

\bibitem{peters2013optical}
Peters-Limbach, M.~A., Groff, T.~D., Kasdin, N.~J., Driscoll, D., Galvin, M.,
  Foster, A., Carr, M.~A., LeClerc, D., Fagan, R., McElwain, M.~W., et~al.,
  ``The optical design of charis: an exoplanet ifs for the subaru telescope,''
  in [{\em Techniques and Instrumentation for Detection of Exoplanets
  VI}{\nolinebreak\hspace{0.1em}]},   {\bf 8864},  88641N, International
  Society for Optics and Photonics (2013).

\bibitem{brandt2014charis}
Brandt, T.~D., McElwain, M.~W., Janson, M., Knapp, G.~R., Mede, K., Limbach,
  M.~A., Groff, T., Burrows, A., Gunn, J.~E., Guyon, O., et~al., ``Charis
  science: performance simulations for the subaru telescope's third-generation
  of exoplanet imaging instrumentation,'' in [{\em Adaptive Optics Systems
  IV}{\nolinebreak\hspace{0.1em}]},   {\bf 9148},  914849, International
  Society for Optics and Photonics (2014).

\bibitem{claudi2008sphere}
Claudi, R.~U., Turatto, M., Gratton, R.~G., Antichi, J., Bonavita, M., Bruno,
  P., Cascone, E., De~Caprio, V., Desidera, S., Giro, E., et~al., ``Sphere ifs:
  the spectro differential imager of the vlt for exoplanets search,'' in [{\em
  Ground-based and Airborne Instrumentation for Astronomy
  II}{\nolinebreak\hspace{0.1em}]},   {\bf 7014},  70143E, International
  Society for Optics and Photonics (2008).

\bibitem{mesa2015performance}
Mesa, D., Gratton, R., Zurlo, A., Vigan, A., Claudi, R., Alberi, M., Antichi,
  J., Baruffolo, A., Beuzit, J.-L., Boccaletti, A., et~al., ``Performance of
  the vlt planet finder sphere-ii. data analysis and results for ifs in
  laboratory,'' {\em Astronomy \& Astrophysics}~{\bf 576},  A121 (2015).

\bibitem{doelman2017patterned}
Doelman, D.~S., Snik, F., Warriner, N.~Z., and Escuti, M.~J., ``Patterned
  liquid-crystal optics for broadband coronagraphy and wavefront sensing,'' in
  [{\em Techniques and Instrumentation for Detection of Exoplanets
  VIII}{\nolinebreak\hspace{0.1em}]},   {\bf 10400},  104000U, International
  Society for Optics and Photonics (2017).

\bibitem{otten2017sky}
Otten, G.~P., Snik, F., Kenworthy, M.~A., Keller, C.~U., Males, J.~R.,
  Morzinski, K.~M., Close, L.~M., Codona, J.~L., Hinz, P.~M., Hornburg, K.~J.,
  et~al., ``On-sky performance analysis of the vector apodizing phase plate
  coronagraph on magao/clio2,'' {\em The Astrophysical Journal}~{\bf 834}(2),
  175 (2017).

\bibitem{haffert2018lexi}
Haffert, S.~Y., Wilby, M.~J., Keller, C.~U., Snellen, I. A.~G., Doelman, D.~S.,
  Por, E.~H., Wardenier, J., and Fagginger~Auer, F., ``On-sky results of
  high-dispersion integral-field spectroscopy and high-contrast imaging with
  the leiden exoplanet instrument(lexi),'' in [{\em Adaptive Optics Systems
  VI}{\nolinebreak\hspace{0.1em}]},   {\bf 10703}, International Society for
  Optics and Photonics (2018).

\bibitem{snellen2015combining}
Snellen, I., de~Kok, R., Birkby, J., Brandl, B., Brogi, M., Keller, C.,
  Kenworthy, M., Schwarz, H., and Stuik, R., ``Combining high-dispersion
  spectroscopy with high contrast imaging: Probing rocky planets around our
  nearest neighbors,'' {\em Astronomy \& Astrophysics}~{\bf 576},  A59 (2015).

\bibitem{beuzit2008sphere}
Beuzit, J.-L., Feldt, M., Dohlen, K., Mouillet, D., Puget, P., Wildi, F., Abe,
  L., Antichi, J., Baruffolo, A., Baudoz, P., et~al., ``Sphere: a planet finder
  instrument for the vlt,'' in [{\em Ground-based and airborne instrumentation
  for astronomy II}{\nolinebreak\hspace{0.1em}]},   {\bf 7014},  701418,
  International Society for Optics and Photonics (2008).

\bibitem{jovanovic2015subaru}
Jovanovic, N., Martinache, F., Guyon, O., Clergeon, C., Singh, G., Kudo, T.,
  Garrel, V., Newman, K., Doughty, D., Lozi, J., et~al., ``The subaru
  coronagraphic extreme adaptive optics system: enabling high-contrast imaging
  on solar-system scales,'' {\em Publications of the Astronomical Society of
  the Pacific}~{\bf 127}(955),  890 (2015).

\bibitem{skrutskie2010large}
Skrutskie, M., Jones, T., Hinz, P., Garnavich, P., Wilson, J., Nelson, M.,
  Solheid, E., Durney, O., Hoffmann, W., Vaitheeswaran, V., et~al., ``The large
  binocular telescope mid-infrared camera (lmircam): final design and status,''
  in [{\em Ground-based and Airborne Instrumentation for Astronomy
  III}{\nolinebreak\hspace{0.1em}]},   {\bf 7735},  77353H, International
  Society for Optics and Photonics (2010).

\bibitem{doelman2018first}
Doelman, D.~S., Por, E.~H., Bos, S.~P., Lozi, J., Guyon, O., Jovanovic, N.,
  Groff, T.~D., Warriner, N.~Z., Escuti, M.~J., and Snik, F., ``First light for
  the vapp on scexao/charis,'' in [{\em Adaptive Optics Systems
  VI}{\nolinebreak\hspace{0.1em}]},   {\bf 10703}, International Society for
  Optics and Photonics (2018).

\bibitem{fusco2006high}
Fusco, T., Rousset, G., Sauvage, J.-F., Petit, C., Beuzit, J.-L., Dohlen, K.,
  Mouillet, D., Charton, J., Nicolle, M., Kasper, M., et~al., ``High-order
  adaptive optics requirements for direct detection of extrasolar planets:
  Application to the sphere instrument,'' {\em Optics Express}~{\bf 14}(17),
  7515--7534 (2006).

\bibitem{dohlen2008infra}
Dohlen, K., Langlois, M., Saisse, M., Hill, L., Origne, A., Jacquet, M.,
  Fabron, C., Blanc, J.-C., Llored, M., Carle, M., et~al., ``The infra-red dual
  imaging and spectrograph for sphere: design and performance,'' in [{\em
  Ground-based and Airborne Instrumentation for Astronomy
  II}{\nolinebreak\hspace{0.1em}]},   {\bf 7014},  70143L, International
  Society for Optics and Photonics (2008).

\bibitem{langlois2014high}
Langlois, M., Dohlen, K., Vigan, A., Zurlo, A., Moutou, C., Schmid, H., Mili,
  J., Beuzit, J.-L., Boccaletti, A., Carle, M., et~al., ``High contrast
  polarimetry in the infrared with sphere on the vlt,'' in [{\em Ground-based
  and Airborne Instrumentation for Astronomy V}{\nolinebreak\hspace{0.1em}]},
  {\bf 9147},  91471R, International Society for Optics and Photonics (2014).

\bibitem{lozi2018scexao}
Lozi, J., Guyon, O., and et~al., ``Scexao, an instrument with a dual purpose:
  perform cutting-edge science and develop new technologies,'' in [{\em
  Adaptive Optics Systems VI}{\nolinebreak\hspace{0.1em}]},   {\bf 10703},
  International Society for Optics and Photonics (2018).

\bibitem{snik2014combining}
Snik, F., Otten, G., Kenworthy, M., Mawet, D., and Escuti, M., ``Combining
  vector-phase coronagraphy with dual-beam polarimetry,'' in [{\em Ground-based
  and Airborne Instrumentation for Astronomy V}{\nolinebreak\hspace{0.1em}]},
  {\bf 9147},  91477U, International Society for Optics and Photonics (2014).

\bibitem{thalmann2008sphere}
Thalmann, C., Schmid, H.~M., Boccaletti, A., Mouillet, D., Dohlen, K.,
  Roelfsema, R., Carbillet, M., Gisler, D., Beuzit, J.-L., Feldt, M., et~al.,
  ``Sphere zimpol: overview and performance simulation,'' in [{\em Ground-based
  and Airborne Instrumentation for Astronomy II}{\nolinebreak\hspace{0.1em}]},
   {\bf 7014},  70143F, International Society for Optics and Photonics (2008).

\bibitem{norris2015vampires}
Norris, B., Schworer, G., Tuthill, P., Jovanovic, N., Guyon, O., Stewart, P.,
  and Martinache, F., ``The vampires instrument: imaging the innermost regions
  of protoplanetary discs with polarimetric interferometry,'' {\em Monthly
  Notices of the Royal Astronomical Society}~{\bf 447}(3),  2894--2906 (2015).

\bibitem{por2018hcipy}
Por, E.~H., Haffert, S.~Y., Radhakrishnan, V.~M., Doelman, D.~S., Van~Kooten,
  M., and Bos, S.~P., ``High contrast imaging for python (hcipy): an
  open-source adaptive optics and coronagraph simulator,'' in [{\em Adaptive
  Optics Systems VI}{\nolinebreak\hspace{0.1em}]},   {\bf 10703}, International
  Society for Optics and Photonics (2018).

\end{thebibliography}
\bibliographystyle{spiebib} 

\end{document}